\documentstyle[aps,multicol,prl]{revtex}

\begin{document}
\begin{title}
{\bf Nonsteady Condensation and Evaporation Waves}
\end{title}
\author{Osamu Inomoto$^{a}$, Shoichi Kai$^{b}$, and Boris A. Malomed$^{c}$}
\address{$^{a}$Institute of Environment Systems, Kyushu University, \\
Fukuoka 812-8581, Japan\\
$^{b}$Department of Applied Physics, Kyushu University, Fukuoka 812-8581,\\
Japan\\
$^{c}$Department of Interdisciplinary Studies, Faculty of Engineering, Tel\\
Aviv University, Tel Aviv 69975, Israel}
\date{\today}
\maketitle 

\begin{abstract}
We study motion of a phase transition (PT) front at a constant temperature
between stable and metastable states in fluids with the Van der Waals
equation of state. We focus on a case of relatively large metastability and
low viscosity, when no steadily moving PT front exists. Simulating the
one-dimensional hydrodynamic equations, we find that the PT front generates
acoustic shocks in forward and backward directions. Through this mechanism,
the nonsteady PT front drops its velocity and eventually stops. The shock
wave may shuttle between the PT front and the system's edge, rarefaction
waves appearing in the shuttle process. If the viscosity is below a certain
threshold, an instability sets in, driving the system into a turbulent state.
\end{abstract}


Motion of phase transition (PT) fronts is the most important phenomenon in
macroscopic physical kinetics \cite{Langer,MerRum}. In most cases, it is
described in the quasi-equilibrium approximation assuming a critical value
of the temperature at the phase interface \cite{Langer}. A distinct class of
phenomena are {\it fast propagating} phase transformations, such as
detonation \cite{Clavin} and waves supported by mechanisms as different as
relaxation in glassy solids \cite{Shklo}, Belousov-Zhabotinsky reaction \cite
{big}, bubble boiling \cite{boiling}, and the Marangoni flow coupled to a
reaction on a fluid surface or chemical transformations coupled to
mechanical stress \cite{Len}. Coupling of the ``chemistry'' to the flow or
elasticity is a necessary ingredient of the fast propagation of a PT wave.

A related problem is the propagation of condensation and evaporation waves
in fluids. In this case, a consistent description is possible, based on the
combination of the corresponding Navier-Stokes (NS) and heat-transport
equations with the equation of state, e.g., the Van der Waals (VdW) one \cite
{old,newest}. Close to the critical point (CP), the VdW equation of a {\it %
spatially inhomogeneous} state takes the following form in terms of the
normalized pressure $p\equiv (P-P_{{\rm CP}})/P_{{\rm CP}}$, temperature $%
\tau \equiv (T_{{\rm CP}}-T)/T_{{\rm CP}}$, and density $\theta \equiv (\rho
-\rho _{{\rm CP}})/\sqrt{\tau }\rho _{{\rm CP}}$ \cite{LL}: 
\begin{equation}
p+4\tau +3\tau ^{3/2}\left( 2\theta -\theta ^{3}/2\right) +G\sqrt{\tau }%
\Delta \theta =0\,,  \label{VdW}
\end{equation}
where $\Delta $ is the Laplacian, and $G$ is an effective surface tension of
the interface separating the vapor ($\theta <0$) and condensed ($\theta >0$)
phases existing below CP (at $\tau >0$).

The hydrodynamic part of the problem is described by a system of the NS and
continuity equations. In the simplest case when the PT front is
one-dimensional (1D), in terms of the rescaled time $t\equiv \tau \sqrt{%
P_{c}/\rho _{c}G}\,\tilde{t}$ ($\tilde{t}$ is the time proper), coordinate $%
z\equiv \sqrt{\tau /G}\,x$, and flow velocity $w\equiv \tau ^{-1}\sqrt{\rho
_{c}/P_{c}}\,v$, these equations are \cite{old} 
\begin{eqnarray}
w_{t}+\sqrt{\tau }ww_{z} &=&6\theta _{z}-(9/2)\theta ^{2}\theta _{z}+\theta
_{zzz}+gw_{zz}\,,  \label{NS} \\
\theta _{t}+\left[ (1+\sqrt{\tau }\theta )w\right] _{z} &=&0,  \label{cont}
\end{eqnarray}
where $g\equiv \left[ (4/3)\eta +\zeta \right] \left( GP_{{\rm CP}}\rho _{%
{\rm CP}}\right) ^{-1/2}$, $\eta $ and $\zeta $ being shear and bulk
viscosities.

To produce a closed system, one should add the heat transfer equation to
these equations. Analysis performed in Ref. \cite{old} demonstrates that a
flat PT wave in a bulk medium is driven by the heat transfer giving rise to
a slowly advancing front. The situation is drastically different in 1D case,
corresponding, e.g., to the fluid inside a capillar, which allows to impose 
{\it isothermal} conditions, $\tau $ {\rm = const}. However, it is not
necessary to conduct the experiment in a capillar; instead, a PT wave may
spontaneously propagate in a thin film sheathing an isothermal wire in a
bulk medium. In fact, a fast PT front of this kind was already observed in
experiment \cite{boiling}.

In the 1D situation, a solution for a PT front moving at a velocity $-V$ can
be sought for as $w=w(\xi )$, $\theta =\theta (\xi )$, with $\xi =z+Vt$.
Eliminating $w$ and assuming $\theta ^{\prime }(\pm \infty )=0$, one arrives
at an equation 
\begin{eqnarray}
\theta ^{\prime \prime }-gV\left[ 1+\sqrt{\tau }\left( \theta _{2}-2\theta
\right) \right] \theta ^{\prime } &=&(3/2)\left( \theta ^{3}-\theta
_{0}^{3}\right)  \nonumber \\
&&-\left[ 6+V^{2}\left( 1+2\sqrt{\tau }\theta _{2}\right) \right] \left(
\theta -\theta _{0}\right) \,,  \label{single}
\end{eqnarray}
where $\theta _{0}\equiv \theta (-\infty )<0$ is the density of the
supercooled vapor invaded by the condensation front, and $\theta _{2}\equiv
\theta (+\infty )>0$ is the density of the stable liquid phase behind it;
recall that the metastable states have $2/\sqrt{3}<|\theta |<2$, while the
regions $|\theta |>2$ and $|\theta |<2/\sqrt{3}$ are absolutely stable and
unstable, respectively.

An exact solution to Eq. (\ref{single}) satisfying the boundary conditions $%
\theta (\pm \infty )=\theta _{2,0}$ was found in \cite{old}, neglecting the
terms $\sim \sqrt{\tau }$. As is known \cite{MerRum}, an equation of this
type selects the velocity as an eigenvalue, $V(\theta _{0})$. A final result
is that there is a pair of stable and unstable branches of $V(\theta _{0})$,
which meet and disappear at a critical density $\theta _{\ast }=-\sqrt{%
(8/3)\left( 9-2g^{2}\right) /\left( 6-g^{2}\right) }$. At this point $%
V(\theta _{0})$ attains its maximum, $V_{{\rm max}}=18g^{2}\left[ \left(
9-2g^{2}\right) \left( 6-g^{2}\right) \right] ^{-1}$.$\,$ The density of the
stable phase behind the front is also found as a part of the solution, $%
\theta _{2}=-\theta _{0}/2+\left[ (2/3)V^{2}+4-(3/4)\theta _{0}^{2}\right]
^{1/2}$. In particular, in the case $g=0$ (inviscid fluid) $\theta _{\ast
}=-2$, which is exactly the border between the stable and metastable states (%
{\it binodal}), hence a steadily moving PT front cannot exist in inviscid
fluids.

Eq. (\ref{single}) also gives rise to traveling-wave solutions of a
different type, which are acoustic shock waves (ASW) without PT. Unlike the
PT wave for which the density $\theta _{2}$ behind the front is determined
by $\theta _{0}$, in the case of ASW the densities $\theta _{i}$ and $\theta
_{f}$ before and after the front may be arbitrary, the velocity being
determined by the boundary conditions. Neglecting the $\sqrt{\tau }$ terms,
it is 
\begin{equation}
V_{{\rm ASW}}^{2}=(3/2)\left( \theta _{i}^{2}+\theta _{f}^{2}+\theta
_{i}\theta _{f}-4\right) \,.  \label{shock}
\end{equation}
In the limit $\theta _{f}-\theta _{i}\rightarrow 0$, this yields the sound
velocity at a given density, $c^{2}=(9/2)\theta ^{2}-6$. The PT front may
move {\em faster }than sound velocity in the metastable phase invaded by the
front. In particular, the velocity $V_{{\rm max}}$ is transonic, provided
that $g>\sqrt{3}$. Thus, the PT wave is related to detonation, whose
characteristic feature is transonic propagation \cite{MerRum,Clavin}.

The critical density $\theta _{\ast }$ belongs to the metastable region if $%
g<2$. In this case, there is a region of the metastable states, $2/\sqrt{3}%
<|\theta _{0}|<|\theta _{\ast }|$, which {\em cannot} be converted into an
absolutely stable state by a steadily moving PT wave \cite{old}. The
objective of the present work is to investigate the propagating PT front in
this region by means of direct simulations of Eqs. (\ref{NS}) and (\ref{cont}%
) (analytical investigation of weakly nonsteady fronts is difficult, as the
corresponding perturbative expansion contains terms secular in $t$ \cite{old}%
).

Before that, we notice that, because of the$\ $term $\sim \theta \theta
^{\prime }$, Eq. (\ref{single}) with $\sqrt{\tau }$ terms kept in\ it does
not belong to the standard type for which the velocity can be found exactly 
\cite{MerRum}. Nevertheless, an ansatz similar to that solving the standard
equation yields an exact result for Eq. (\ref{single}) as well: $V=-(3\sqrt{3%
}/2g)\theta _{1}\left[ (1+\sqrt{\tau }\theta _{2})\left( 1+\sqrt{\tau }%
(3\theta _{1}+\theta _{2})\right) \right] ^{-1/2}$, $\theta _{1}\neq $ $%
\theta _{0,2}$ being another root of r.h.s. of Eq. (\ref{single}). As $\sqrt{%
\tau }\ll 1$, we look for lowest-order corrections to the final results.
They contain a new effect, breaking the symmetry between the condensation ($%
+ $) and evaporation ($-$) waves: the $\sqrt{\tau }$-corrected threshold
value of the viscosity $g$, below which there is a region of the
nonexistence of steady PT waves, is $2\pm (5/2)\sqrt{\tau /3}$.

In systematic numerical simulations of Eqs. (\ref{NS}) and (\ref{cont}), the 
$\sqrt{\tau }$-corrections were omitted, and $w$ was eliminated,
differentiating Eq. (\ref{NS}) in $z$ and substituting $w_{z}$ by $-\theta
_{t}$ as per Eq. (\ref{cont}) (for some typical cases, it was checked that
the $\sqrt{\tau }$ terms produce a small insignificant change of the
numerical solution). The resulting equation of the second order in $t$ and
fourth order in $z$ was numerically integrated with the initial conditions $%
\theta _{t}(t=0)=0$, $\theta (t=0,z)$ being a smoothed step between the
values $\theta _{0}<0$ and $\theta _{2}>0$. For the boundary conditions
(b.c.), it was always taken that $\theta =\theta _{0}$ at the left edge and $%
\theta _{z}=0$ at both edges. Note that the natural b.c. for the flow, $%
w(-\infty )=0$, is compatible with these, as is seen from Eq. (\ref{cont}).
The b.c. for the flow at the right edge is not fixed, because, as a matter
of fact, we are dealing with an open system, with a possible influx of fluid
from $+\infty $.

In the case when the steady PT front exists, the fourth b.c. was naturally
taken as $\theta (+\infty )=\theta _{0}$. Simulations have demonstrated that
the steady front is always stable when it exists. When it does not exist,
the density to be established past the PT wave is unknown, therefore initial 
$\theta _{2}$ was taken arbitrarily from the absolute stability region $%
|\theta |>2$. In this case, the missing b.c. was taken as $\theta _{zz}=0$
(i.e., the density profile must be flat) at the right edge. The results
displayed below clearly show that no artifacts are generated by b.c.

A representative picture of the {\it nonsteady} propagation of the
condensation front, when the steady one does not exist, is shown in Fig. 1.
Essential features revealed by many simulations are well seen in it: (i) the
nonsteady PT wave immediately begins to generate two ASWs in forward and
backward directions; (ii) the forward ASW ({\it precursor}) travels
essentially faster than the PT front, so that a rapidly expanding trough
with an increased value $|\theta |>|\theta _{0}|$ is formed ahead of the
front; (iii) the fast backward-propagating ASW hits the right edge, bounces
from it, and then hits the PT front. Velocities of all the observed ASWs
were checked to comply with Eq. (\ref{shock}).

Thus, the first finding is that the nonsteady PT wave does not
self-accelerate, which could be natural to expect \cite{old}; instead, it
generates a strong acoustic precursor. Fig. 2a demonstrates that in all the
cases with $g\leq \sqrt{3}$ the newly established value $\tilde{\theta}$ of
the density ahead of the front originally turns out to be fairly close to $%
\theta _{\ast }$, corresponding to the minimum of $|\theta _{0}|$ at which
the steady PT wave exists. Then, quasi-steady propagation of the PT wave
with the density $\tilde{\theta}\approx \theta _{\ast }$ ahead of it becomes
possible. Note that simulations of 2D hydrodynamic equations with the VdW
equation of state, in the problem of transverse instability of shock waves,
have also revealed strong emission of acoustic waves \cite{newest}.

The PT front decelerates and eventually comes to halt, while the precursor
keeps propagating at a nearly constant velocity until it hits the left edge
of the integration domain (Fig. 2b). Moreover, simulations show that the PT
front begins then to slowly move in the reverse direction. We do not
consider the reverse motion in detail, as it is a manifestation of
finiteness of the system, while in this work we focus on properties of a
semiinfinite (to the left) system, although, of course, finiteness effects
may be important for experiment. As concerns the value of $\left| \theta
\right| $ ahead of the PT front, at a later stage of the evolution it slowly
increases from the originally established value $\approx |\theta _{\ast }|$
to $|\theta |=2$ at the halt stage. This can be easily understood, as $%
|\theta |=2$ is the {\em binodal} (a minimum value of $|\theta |$ at which
the interface between the two phases may be quiescent). The halt of the
front at $\tilde{\theta}=-2$ implies that the density behind it must take
the conjugate value $\theta =+2$, which is indeed observed, see below.

An intriguing case is $\sqrt{3}<g<2$, when the nonsteady situation is
possible whilst the maximum velocity of the steadily moving PT exceeds the
sound velocity at $\theta =\theta _{\ast }$, suggesting that the PT front
cannot send a sound wave ahead of itself. A typical picture for this case is
displayed in Fig. 3, which shows that the acoustic precursor is nevertheless
launched. The difference from the previous case is that the trough before
the PT front quickly deepens, so that $\tilde{\theta}$, rather than being
stuck at the value $\theta _{\ast }$, quickly drops to the binodal value $%
\theta =-2$, see Fig. 2a. This leads to PT coming to halt earlier than in
the case $g\leq \sqrt{3}$. Thus, an attempt to launch a ``superfast'' PT
front into a deeply metastable phase produces an opposite result: the front
generates strong acoustic waves and decelerates, seeing a less metastable
state ahead of itself (physically, the metastability is relatively deep, as
everything happens close to the critical point).

An altogether different result is found at very small $g$, viz., onset of
instability, which quickly switches the system into a fully turbulent state
and destroys the interface. For instance, with the initial values $\theta
_{0}=-1.2$ before the interface and $\theta =2.7$ behind it, the instability
sets in at $g\leq 0.3$. However, systematic study the present system as a
model of 1D turbulence is beyond the scope of this work.

The situation behind the PT front deserves special description. We observe
that ASW which hits the PT front in the last configuration shown in the
upper part of Fig. 1 bounces from the front, travels to the right edge, is
reflected from it, and then again hits the PT front (Fig. 4). A nontrivial
feature of ASW in this case, evident in Fig. 4, is that it is reflected by
the PT front as an {\it antishock}, or rarefaction wave. The present model
does not admit steady rarefaction waves; however, the wave in Fig. 4 is
unsteady, as it propagates between variable densities. Taking instantaneous
values of the densities, we conclude that the antishock's velocity also
obeys Eq. (\ref{shock}).

Analysis of the numerical results leads to the following inferences
concerning the shuttling shock/antishock wave behind the PT front in Figs. 1
and 4: (i) bounces from both the right edge of the system and PT front do
not change the absolute value of the velocity; (ii) bounces from the right
edge are elastic without a change in the step height ($\theta _{f}-\theta
_{i}$); (iii) bounces from the PT front are highly {\em inelastic}: the
shock bounces as an antishock and vice versa, each reflection decreasing the
step height by a factor $\sim 5$, which explains why the shuttling wave
disappears after a few reflections.

Thus, we conclude that the system eventually drives itself into an
equilibrium state with a quiescent interface between the liquid and vapor
phases. A mechanism establishing the equilibrium is the generation of
acoustic shocks, which bounce elastically from the system edge, but are
muffled by strongly inelastic collisions with the interface.

\newpage

\section*{Figure Captions}

Fig. 1. Propagation of the nonsteady condensation front with $g=\sqrt{3}$.
This and other figures display results for a typical case with the initial
densities before and behind the front $\theta _{0}=-1.2$ and $\theta
_{2}=+2.7$ (note that $\theta _{\ast }(g=\sqrt{3})=-1.63$). Profiles $\theta
(z)$ are shown through time intervals $\Delta t$, different in the upper and
lower portions. An interval $-1<\theta <2.5$, where nothing happens, is
dropped. The bold arrows in the upper portion (and in Fig. 4) show
velocities of the shuttling acoustic shock behind the front.

Fig. 2. (a) The difference between the density $\tilde{\theta}$, established
ahead of the phase transition front after the passage of the acoustic
precursor, and the critical density $\theta _{\ast }$. (b) The velocities of
the acoustic shock and condensation front for $g^{2}=2.4,\,3.0$, and $3.6$.
A small jump at $t\sim 200$ is due to hitting the front by the acoustic
shock reflected from the right edge.

Fig. 3. Evolution of the nonsteady condensation front at $g^2=3.6$.

Fig. 4. The propagation of the antishock at $g=\sqrt{3}$ between two
inelastic collisions with the condensation front: before (a) and after (b)
elastic reflection from the right edge.


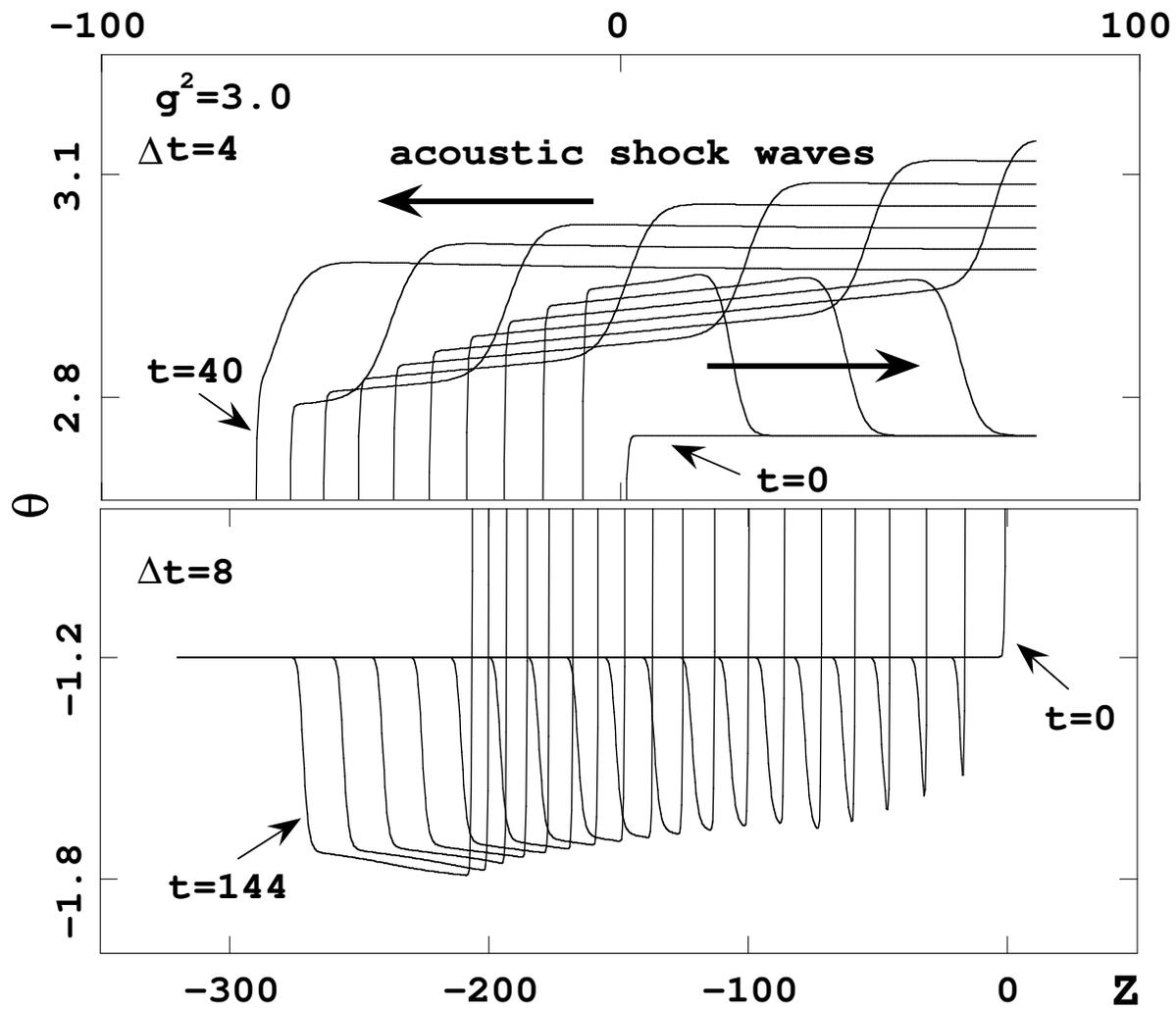
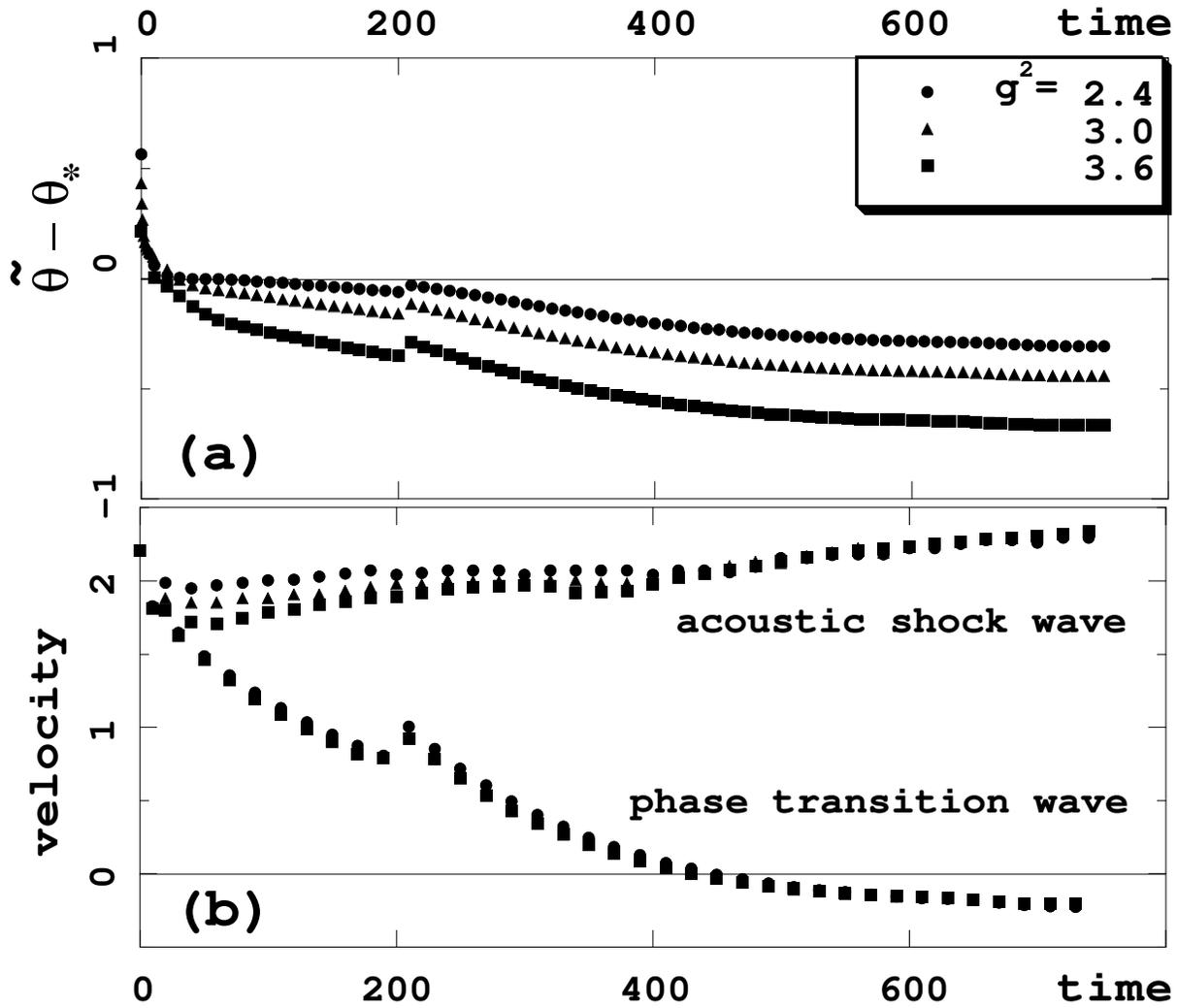
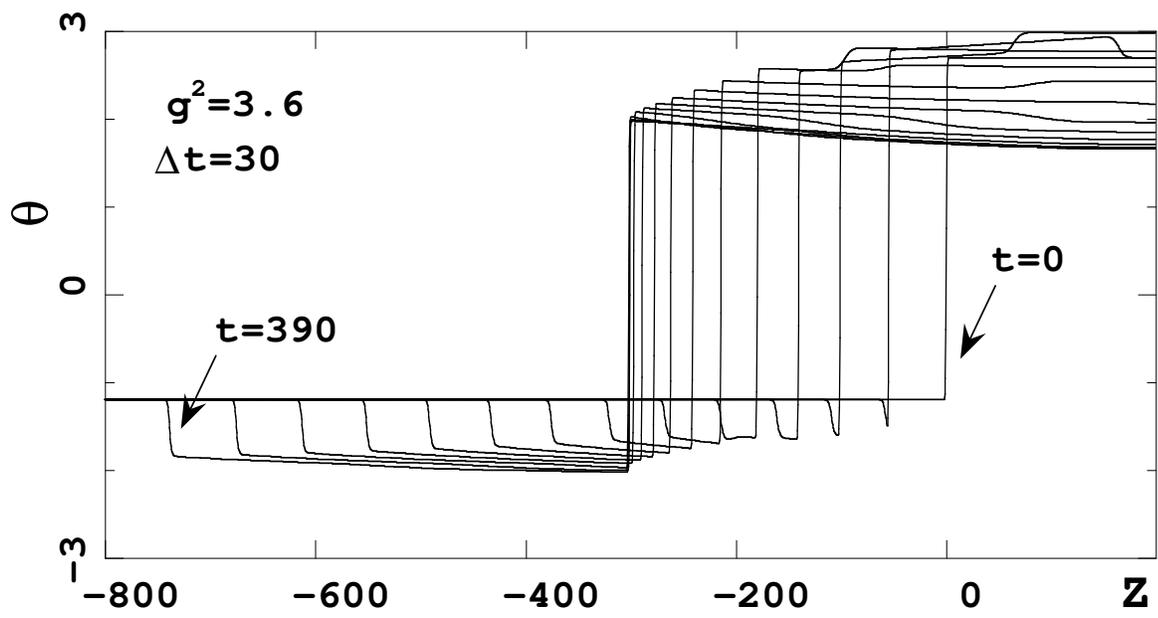
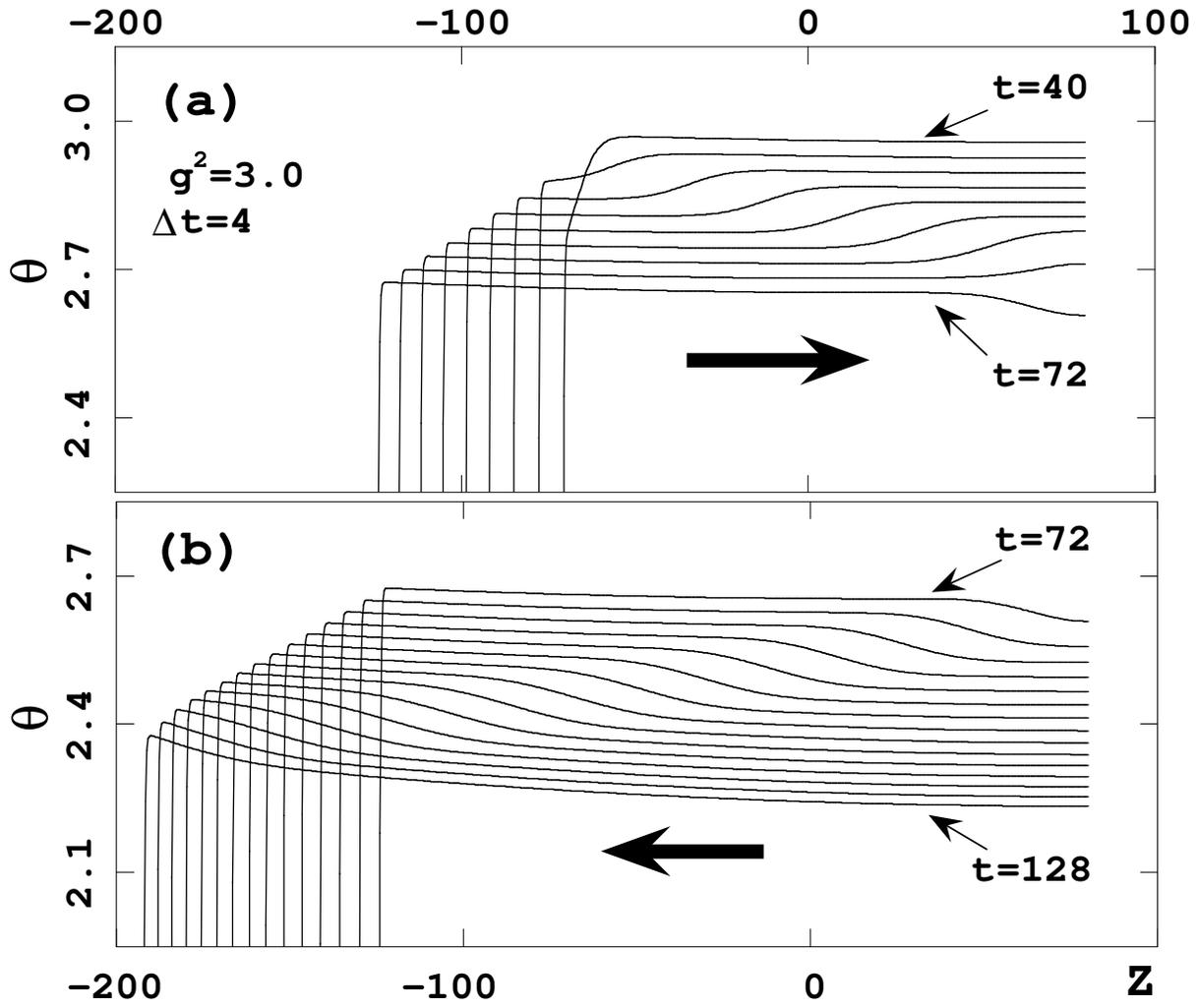
\end{document}